\documentclass[aps,prl,reprint,superscriptaddress,amssymb,citeautoscript]{revtex4-2}
\usepackage{comment}
\usepackage{amsmath}
\usepackage{xcolor}
\usepackage{natbib}
\usepackage{graphicx}
\usepackage{hyperref}
\usepackage[normalem]{ulem} 
\hypersetup{
    colorlinks=true,
    linkcolor=blue,
    filecolor=cyan,
    urlcolor=magenta,
}

\begin{document}

\title{La substitution studies on the heavy-fermion superconductor CeRh$_2$As$_2$}

\author{Sushma Lakshmi Ravi Sankar}
\affiliation{Max Planck Institute for Chemical Physics of Solids, N\"othnitzer Stra{\ss}e 40, 01187 Dresden, Germany}
\affiliation{Institute for Solid State and Materials Physics, Technische Universit\"at Dresden, 01062 Dresden, Germany}

\author{Manuel Brando}
\affiliation{Max Planck Institute for Chemical Physics of Solids, N\"othnitzer Stra{\ss}e 40, 01187 Dresden, Germany}

\author{Jochen Wosnitza}
\affiliation{Institute for Solid State and Materials Physics, Technische Universit\"at Dresden, 01062 Dresden, Germany}
\affiliation{Dresden High Magnetic Field Laboratory (HLD-EMFL) and W\"urzburg-Dresden Cluster of Excellence ctd.qmat, Helmholtz-Zentrum Dresden-Rossendorf, 01328 Dresden, Germany}

\author{Seunghyun Khim}
\email[Corresponding author: ]{Seunghyun.Khim@cpfs.mpg.de}
\affiliation{Max Planck Institute for Chemical Physics of Solids, N\"othnitzer Stra{\ss}e 40, 01187 Dresden, Germany}

\date{\today}

\begin{abstract}
CeRh$_2$As$_2$ has been receiving considerable attention due to its unusual two-phase superconductivity.
The superconducting (SC) phase appears at $T_{\mathrm{c}}$ = 0.35 K in an ordered state (phase I) of the Ce-4$f$ moments, which develops below $T_{\mathrm{0}}$ = 0.55 K.
The microscopic nature of phase I has not been fully established yet.
We report a single-crystal study of the effect of La substitution on these low-temperature phases in Ce$_{1-\textit{x}}$La$_\textit{x}$Rh$_2$As$_2$ up to $\textit{x}$ = 0.1.
The lattice parameters increase monotonically with $x$, corresponding to an effective negative pressure of approximately -0.3 GPa for \textit{x} = 0.1.
While the Ce$^{3+}$ local valence state and the non-Fermi-liquid behavior are preserved, the resistivity coherence maximum $T^{*}_{\mathrm{max}}$ $\approx$ 45 K in the pristine sample shifts to lower temperatures with increasing $x$, indicating a suppression of the Kondo energy scale upon lattice expansion.
On the other hand, both $T_{\mathrm{c}}$ and $T_{\mathrm{0}}$ are rapidly suppressed to below 0.1 K for \textit{x} $>$ 0.05.
Notably, such a rapid decrease of $T_{\mathrm{c}}$ under moderate negative pressure is unexpected, but is rather consistent with substitution-induced disorder effect which leads to strong pair breaking in unconventional superconductors. The observed fragility of phase I under both pressure and disorder may imply its itinerant origin.

\end{abstract}

\maketitle

\section{Introduction}

CeRh$_2$As$_2$ is an intriguing heavy-fermion superconductor. 
Although its superconducting transition temperature is comparatively low, $T_\mathrm{c}$ = 0.35 K, the upper critical field ($H_{\mathrm{c2}}$) for field aligned along the \textit{c} axis is as high as 18 T \cite{khim2021}.
This $H_{\mathrm{c2}}$ significantly exceeds the Pauli-limiting field of $H_P$ = 1.84 $T_{\mathrm{c}}$ $\approx$ 0.6 T, which sets an upper limit for weak-coupling spin-singlet superconductivity, and, thus, strongly hints at possible spin-triplet pairing.
Indeed, the SC phase diagram demonstrates a phase boundary within the SC state indicating a transition from a low-field even-parity (SC1) state to a high-field odd-parity (SC2) state \cite{Landaeta2022}.
This behavior has been attributed to the cooperation of heavy-fermion superconductivity and strong spin-orbit coupling that originates from the lack of local inversion symmetry at the Ce site in the centrosymmetric crystal structure \cite{Madar1987,Yoshida2012,Maruyama2012,Sigrist2014,Schertenleib2021,Mockli2021,cavanagh2022,Fischer2023}.
Furthermore, the high-field SC2 phase is proposed to have non-trivial topology \cite{Nogaki2021,Ishizuka2024}.

In addition, the Ce-4$f$ moments order at $T_{\mathrm{0}}$ = 0.55~K, forming a state, called phase I, whose microscopic nature has not been fully established yet \cite{Chajewski2024}.
While antiferromagnetism has been suggested by the presence of a spontaneous internal field detected below $T_{\mathrm{0}}$ \cite{Kibune2021,Ogata2023,khim2025}, this picture is challenged by experimental observations: an in-plane magnetic field enhances $T_{\mathrm{0}}$, and further induces a cascades of new phases \cite{Hafner2022,Semeniuk2023,blawat2024,khanenko2025}.  
An alternative explanation involves higher multipolar-order of the Ce-4$f$ moments.
A quasi-quartet crystal-electric field (CEF) ground state \cite{Christovam2024} combined with a Kondo energy scale comparable to the CEF splitting can stabilize quadrupole degrees of freedom despite this is usually forbidden in tetragonal systems.
Accordingly, several theoretical models have been proposed based on a local picture with cooperating dipole and multipolar orders \cite{Schmidt2024,Thalmeier2025} and an itinerant scenario involving Fermi-surface nesting \cite{Hafner2022}.

The SC and phase I might be intimately linked in terms of quantum criticality.
Non-thermal fluctuations, which arise as the ordered state is suppressed toward zero temperature, possibly mediate unconventional SC pairing.
This is supported by a departure from Fermi-liquid behavior in both specific heat and resistivity, signaling strong electronic correlations and proximity to a quantum critical point (QCP) \cite{khanenko2024}.
Thus, identifying the microscopic nature of the $T_{\mathrm{0}}$ order will help illuminating the SC pairing mechanism as well as the origin of the multiple SC phases in CeRh$_2$As$_2$ \cite{amin2024,szabo2024,lee2025}.

Strong electronic correlations, as evidenced by the quantum-critical behavior and the small energy scales associated with the ordered phases, imply that the fundamental properties of CeRh$_2$As$_2$ are highly sensitive to external parameters.
By systematically tuning the external parameters such as pressure, magnetic field, and compositions, one can associate physical quantities with underlying microscopic interactions.
Recent studies demonstrated a tunability of the SC and phase I under hydrostatic pressures \cite{Siddiquee2023,Semeniuk2024,Pfeiffer2024}.
A QCP was revealed at $P_c$ = 0.5 GPa, where $T_{\mathrm{0}}$ is almost suppressed, $T_{\mathrm{c}}$ is maximized, and a $T$-linear resistivity becomes most pronounced \cite{Pfeiffer2024}.
This indicates an intimate connection between superconductivity and the ordering of the Ce-4$f$ moments.

In the present work, we have characterized La-substituted CeRh$_2$As$_2$ single crystals to probe the nature of the low-temperature phases through  non-magnetic substitution on the Ce site.      
While La substitution up to 10 \% preserves the local Ce$^{3+}$ valence state and the non-Fermi-liquid behavior observed in the pristine compound, it induces a monotonic expansion of the unit-cell volume.
This lattice expansion, corresponding to an effective negative pressure, likely shifts the local resistivity maximum toward lower temperatures by reducing the Kondo energy scale. 
However, the rapid suppression of both $T_{\mathrm{c}}$ and $T_{\mathrm{0}}$ with increasing La concentration suggest that the concomitant disorder effect dominates over the pressure effect.
We discuss how these observations provide insights into the unconventional SC gap structure and the itinerant nature of phase I.

\section{Experiment}
We have grown single crystals of Ce$_{1-\textit{x}}$La$_\textit{x}$Rh$_2$As$_2$ ($x$ = 0.01, 0.03, 0.04, 0.05, and 0.1, where $\mathit{x}$ denotes the nominal concentration of La) using the bismuth (Bi) flux-growth method \cite{khim2021,Semeniuk2023}.
Powder x-ray diffraction (PXRD) measurements were used to confirm the phase purity and to characterize the crystal structure.
To endure accurate determination of the lattice parameters, LaB$_6$ was used as a standard \cite{wincsd}.
The chemical compositions of the crystals were determined by using energy dispersive x-ray spectroscopy (EDX). 
The measurements were performed on at least five distinct spots across the sample surface, which confirms uniform distribution of the elemental composition.
Magnetization and specific-heat measurements were performed in a Quantum Design MPMS (Magnetic Property Measurement System) and PPMS (Physical Property Measurement System), respectively.  
Electrical resistivity was measured using the standard four-probe method in the PPMS.
Low-temperature resistivity characterizations below 1.8 K were obtained by using a customized adiabatic demagnetization refrigeration probe in the PPMS. 

\section{\label{sec:level1}Results}
\subsection{\label{sec:level2}Crystal structure and chemical composition}

PXRD measurements combined with Rietveld refinement confirmed that the Ce$_{1-x}$La$_x$Rh$_2$As$_2$ single crystals crystallize in the CaBe$_2$Ge$_2$-type structure, identical to that of the pristine compound \cite{Madar1987}.  
No additional diffraction peaks were observed, indicating the absence of secondary phases. 
The actual La composition determined by EDX ($x_\textrm{EDX}$) is systematically higher than the nominal value $x$  by a factor of approximately 1.4 ($x_\textrm{EDX}$ = 0.0404 $\pm$ 0.0035, 0.0633 $\pm$ 0.0024, and 0.145 $\pm$ 0.0022 for $x$ = 0.03, 0.5, and 0.1, respectively).
For \textit{x} $<$ 0.03, the La concentration is below the instrumental detection limit of EDX. 

\begin{figure}[h]
    \centering
    \includegraphics[width=1\linewidth]{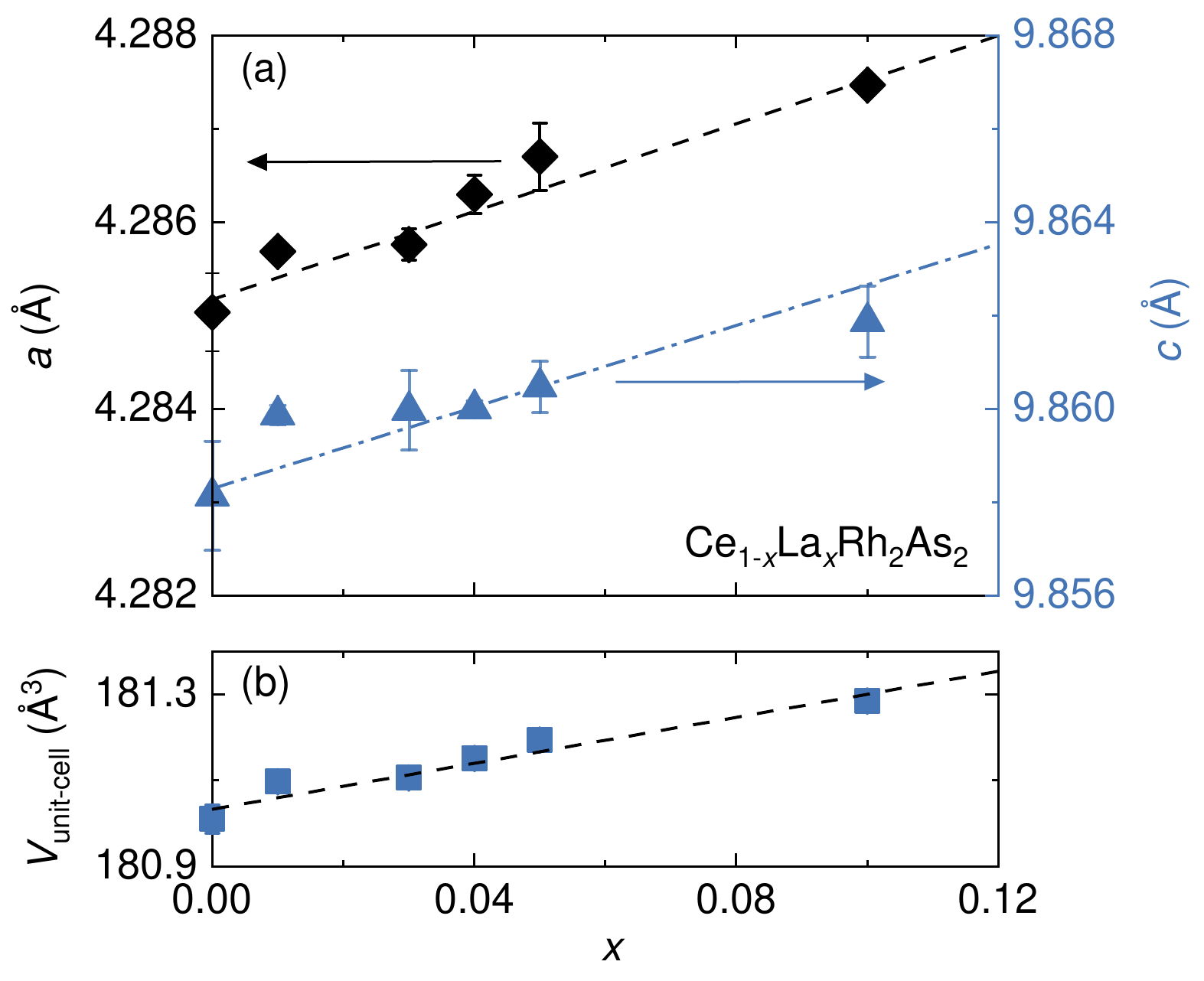}
    \caption{La-concentration (\textit{x}) dependence of (a) the tetragonal unit-cell lattice parameters, \textit{a} and \textit{c}, and (b) the calculated unit-cell volume, $V_{\mathrm{unit-cell}}$.}
    \label{fig:PXRDlattice}
\end{figure}

The lattice parameters determined using the LaB$_6$ standard are shown in Fig.~\ref{fig:PXRDlattice}.
Both the $a$ and $c$ lattice parameters monotonically increase with increasing $x$, following Vegard's law.
This behavior is consistent with the larger atomic radius of La (187 pm) compared to Ce (181 pm), leading to a systematic lattice expansion upon substitution.
For $x$ = 0.1, the $a$ and $c$ lattice parameter increase by 0.057 \% and 0.038 \%, respectively, resulting in a total unit-cell volume increase of approximately 0.15 \%. 
An increase of the unit-cell volume by La substitution has also been observed in other Ce-based heavy-fermion superconductors, such as CePt$_2$Si$_2$ \cite{Ragel2008}, CeCu$_2$Si$_2$ \cite{onuki1987,ANDRAKA1991,ocko2001}, and CeCoIn$_5$ \cite{Petrovic2002}.
In contrast to the relatively isotropic expansion observed in CeRh$_2$As$_2$, the isomorphic (Ce,La)Cu$_2$Si$_2$ system in the ThCr$_2$Si$_2$-type structure exhibits a pronounced increase in the $a$ parameter with minimal change in the $c$ parameter \cite{ocko2001}.
The CaBe$_2$Ge$_2$-type structure features a strong bonding along the $c$ axis, particularly between Rh(2) and As(2), with a Ce atom positioned intermediately \cite{Ragel2008,landaeta2022_prb}.
The substitution of Ce with La will directly disturb the Rh(2)-As(2) bonding, rendering a more significant expansion along the $c$ axis.

\subsection{Magnetization}
\begin{figure}[h]
    \centering
    \includegraphics[width=1\linewidth]{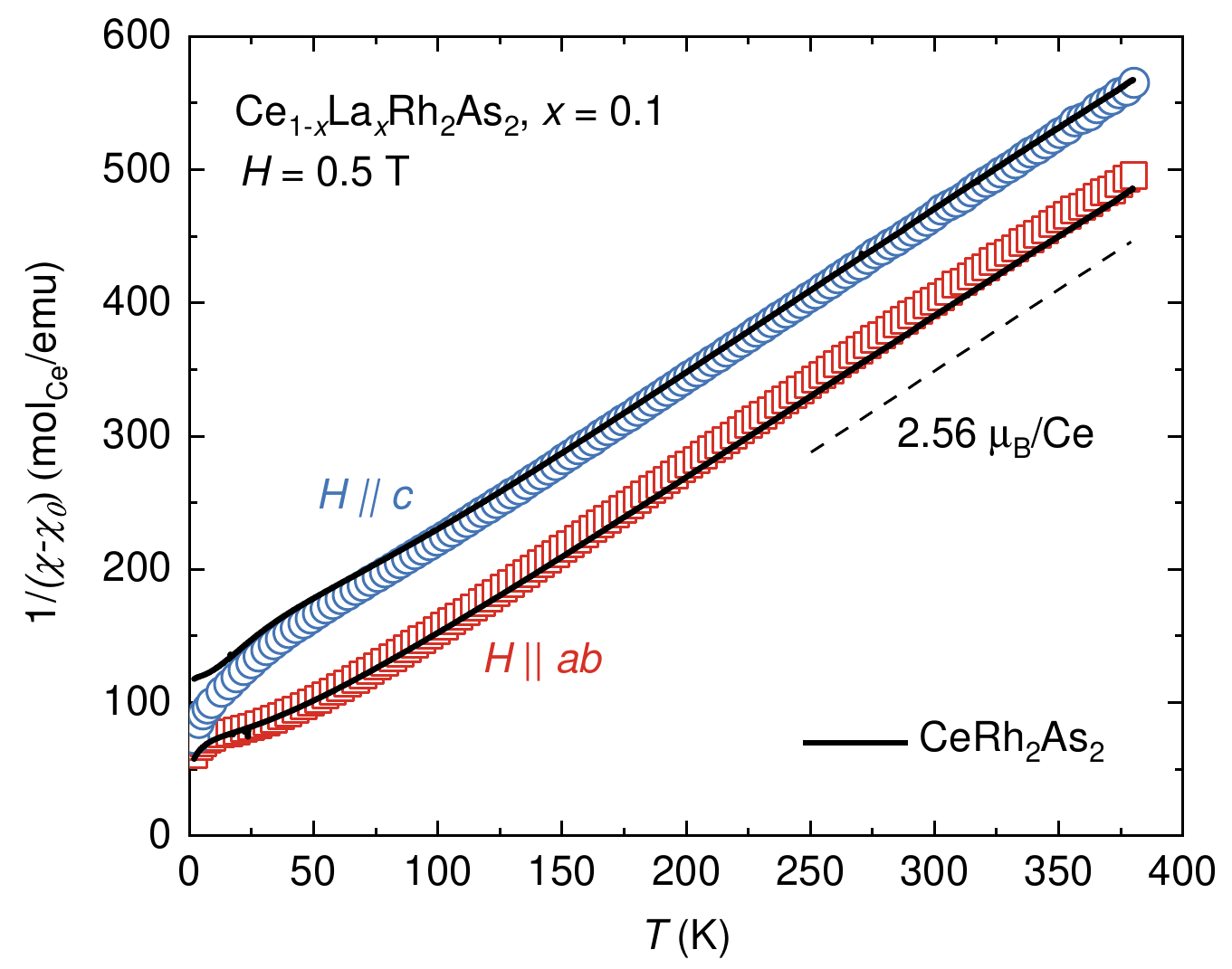}
    \caption{Temperature dependence of the inverse magnetic susceptibility ($\chi$) after subtracting a temperature-independent term ($\chi_0$) for in-plane ($H$ $\parallel$ $ab$, red open squares) and $c$-axis ($H$ $\parallel$ $c$, blue open circles) magnetic field, respectively. The magnetic susceptibility per Ce is obtained by normalization using the $x_\mathrm{EDX}$ value. The solid lines denote the magnetic-susceptibility data of pristine CeRh$_2$As$_2$, reported earlier \cite{khim2021}. 
    The slope of the linear dashed line stands for 2.56 $\mu_B$/Ce, which is the expected effective moment of the Ce$^{3+}$ valence state.}
    \label{fig:chi_La10}
\end{figure}

Figure \ref{fig:chi_La10} depicts the temperature-dependent inverse magnetic susceptibility data of the $x$ = 0.1 sample, the highest La concentration studied. 
To access the impact of the La substitution on the local magnetic state of Ce, we focus on this sample representing the most significant chemical perturbation in our compounds studied.
Above 100 K, the linear slopes of the Curie-Weiss behavior are seen for both field directions.
The effective moment, determined by using the high-temperature fits, is close to 
2.56 ${\mu}_B$/Ce for the Ce$^{3+}$ valence state.
No significant change of $\chi$ is observed in the \textit{x} = 0.1 sample compared to the pristine CeRh$_2$As$_2$.
From the high-temperature Curie-Weiss fit, the Curie-temperature was obtained to be $\theta_{ab}$ = 24.8 K and $\theta_{c}$ = 82.8 K. The estimated $B_{2}^{0}$ parameter in the Hamiltonian describing the CEF model is 6.04 K. 
This value is slightly smaller than  $B_{2}^{0}$  = 6.5 K of the pristine CeRh$_2$As$_2$ \cite{khim2021}. However, by applying the same value of $B_{4}^{0}$ = 0.1 K and $B_{4}^{4}$ = 2.8 K used for CeRh$_2$As$_2$, we could also obtain a fairly good fit (see Supplemental Material).
This supports the assumption that the local Ce state is intact up to $x$ = 0.1, allowing us to consider the substitution to be perturbative.
A more pronounced Curie-Weiss tail is observed for $H$ $\parallel$ \textit{c} (blue open circles in Fig. \ref{fig:chi_La10}) at low temperatures, appearing as a sharper drop in 1/$\chi$.
This anomaly may indicate the presence of paramagnetic impurities, possibly introduced by the chemical substitution \cite{SM}.

\subsection{Specific heat}
\begin{figure}[t]
    \centering
    \includegraphics[width=1\linewidth]{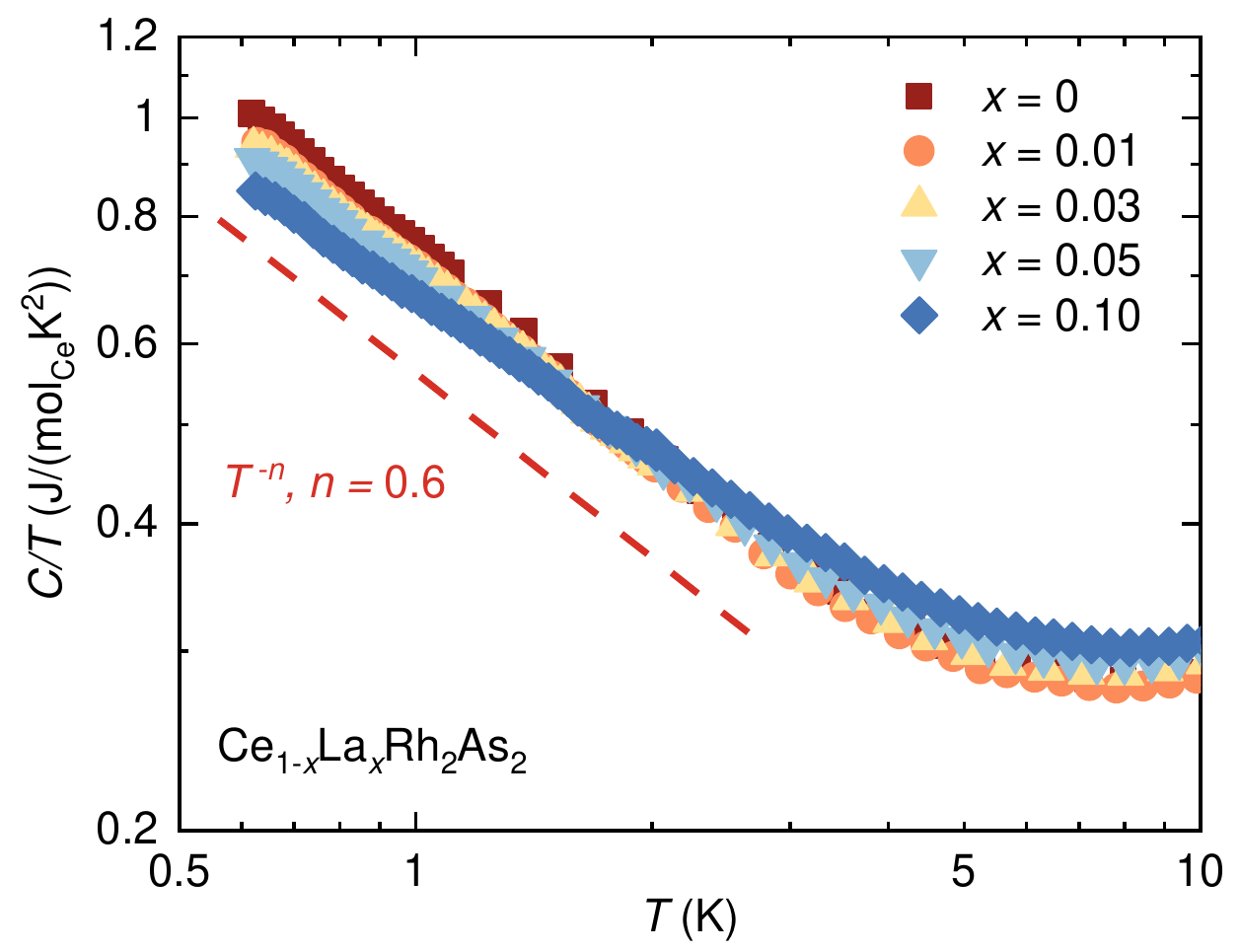}
    \caption{Specific heat divided by temperature for different La concentrations, $x$. \textit{C/T} was normalized by the Ce concentration determined by EDX (1-$x_\mathrm{edx}$).
    }
    \label{fig:CTT}
\end{figure}

Figure~$\ref{fig:CTT}$ shows the temperature dependence of the specific heat divided by temperature, $C$/$T$, for various $x$.
Since the La substitution is expected to minimally affect the lattice-phonon contribution, the observed evolution of $C$/$T$ with $x$ primarily reflects changes in the electronic and magnetic specific heat. 
The power-law increase of  $C$/$T$ $\propto$ $T^{n}$ upon cooling persists even at $x$ = 0.1, indicating the continued presence of quantum critical fluctuations.
However the $C$/$T$ value at low temperatures gradually decreases with increasing $x$, and the exponent $n$ is systematically reduced.
This trend reflects a weakening of electronic correlations and a suppression of quantum critical fluctuations.
The reduced $C$/$T$ below 2 K is compensated by a concomitant increase in $C$/$T$ at higher temperatures, suggesting a redistribution of the magnetic entropy while maintaining the quasi-quartet CEF ground state \cite{khim2021,Hafner2022,Christovam2024}.
Further quantitative analysis remains to be done in specific-heat characterizations at even lower temperatures.

\subsection{Resistivity}

\begin{figure}[h]
    \centering
    \includegraphics[width=1\linewidth]{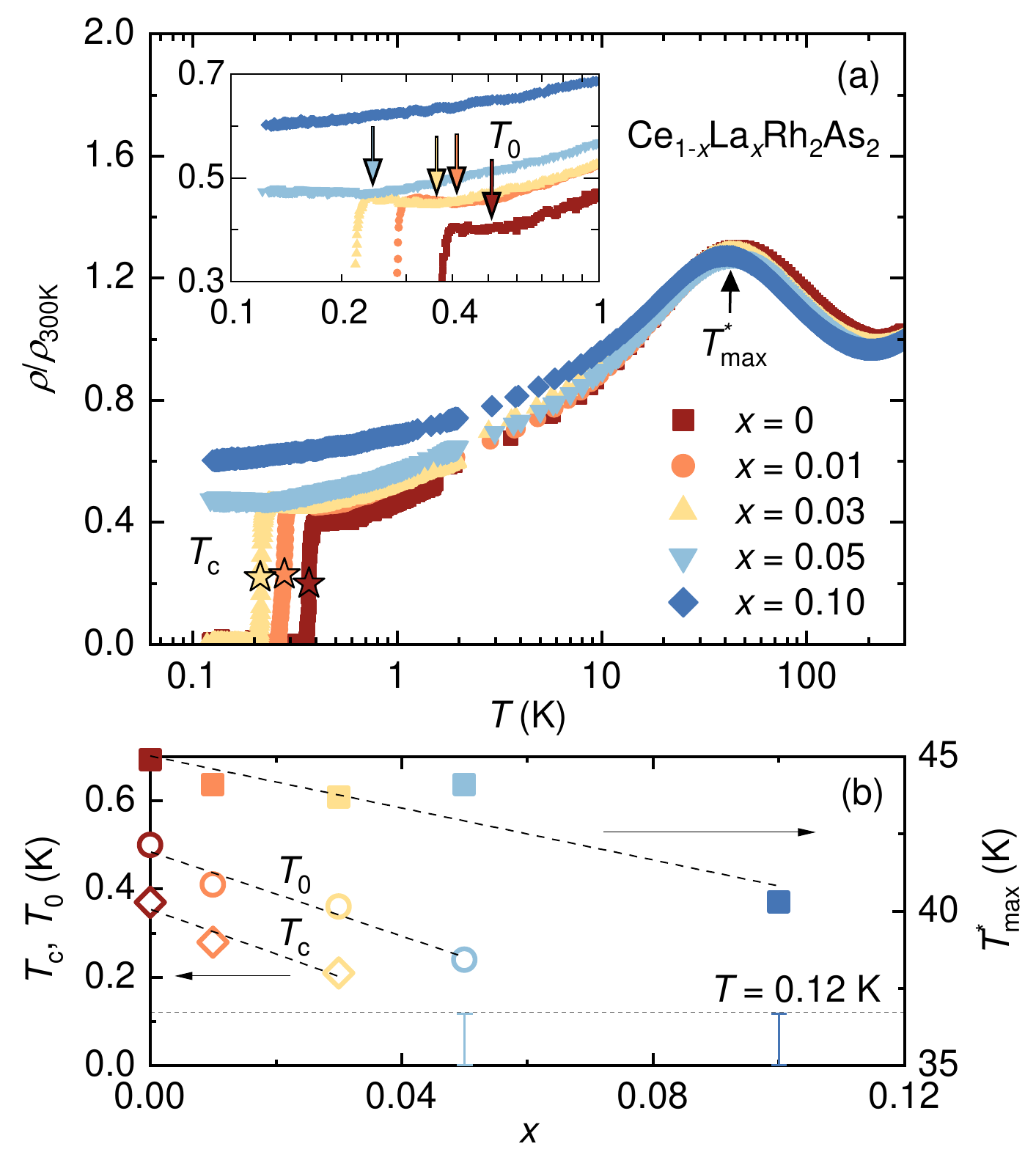}
    \caption{(a) Temperature dependence of the normalized resistivity, $\rho$/$\rho_{\mathrm{300 K}}$ for Ce$_{1-x}$La$_x$Rh$_2$As$_2$. The $T_\mathrm{c}$ and $T_\mathrm{0}$ transition temperatures are marked in the main panel and in the inset, respectively. $T_\mathrm{c}$ is determined as the midpoint of the zero-resistivity transition. $T_\mathrm{0}$ is identified as the temperature at the onset of the resistivity upturn, which correspond to the maximum in the second derivative of the resistivity. Detailed description is shown in the supplemental material \cite{SM}. (b) $T_\mathrm{0}$ (open circles), $T_\mathrm{c}$ (open diamonds), and $T^{*}_{\textrm{max}}$ (solid squares) as a function of nominal La composition \textit{x}. The light blue and dark blue error bars indicate the possibility of $T_c$ and $T_0$ transition temperatures, respectively, below 0.12 K.}
    \label{fig:RTnorm}
\end{figure}

Figure~\ref{fig:RTnorm} displays the normalized temperature-dependent resistivity, $\rho$($T$)/$\rho_{300 K}$.
We note that the room-temperature resistivity value ($\rho_{300 K}$) does not monotonically increase with $x$, probably due to extrinsic factors such as cracks within the sample or uncertainties in determining the sample geometry.
While the normalized resistivities above 50 K are almost identical across all $x$, $\rho$/$\rho_{300 K}$ systematically increases with $x$ below the local maximum at $T^{*}_{\mathrm{max}}$, evidencing increased disorder and disruption of the Kondo coherence.
As a result, $T^{*}_{\mathrm{max}}$ decreases from 45 K for \textit{x} = 0 to 40 K for $x$ = 0.1 [Fig. \ref{fig:RTnorm}(b)].

At lower temperatures, both $T_\mathrm{c}$ and $T_\mathrm{0}$ are reduced with increasing $x$. 
Above $x$ = 0.03, the zero-resistivity transition becomes undetectable, as $T_\mathrm{c}$ is suppressed below 0.12 K, the lowest temperature accessible in our experiments.
Similarly, $T_\mathrm{0}$ shifts to lower temperatures and disappears for $x$ $>$ 0.05.
The suppression rates, given by the slopes of the $T_\mathrm{c}$ and $T_\mathrm{0}$ changes, are -5.9 K/$x$ and -4.8 K/$x$, respectively. 

\begin{figure}[b]
    \centering
    \includegraphics[width=1\linewidth]{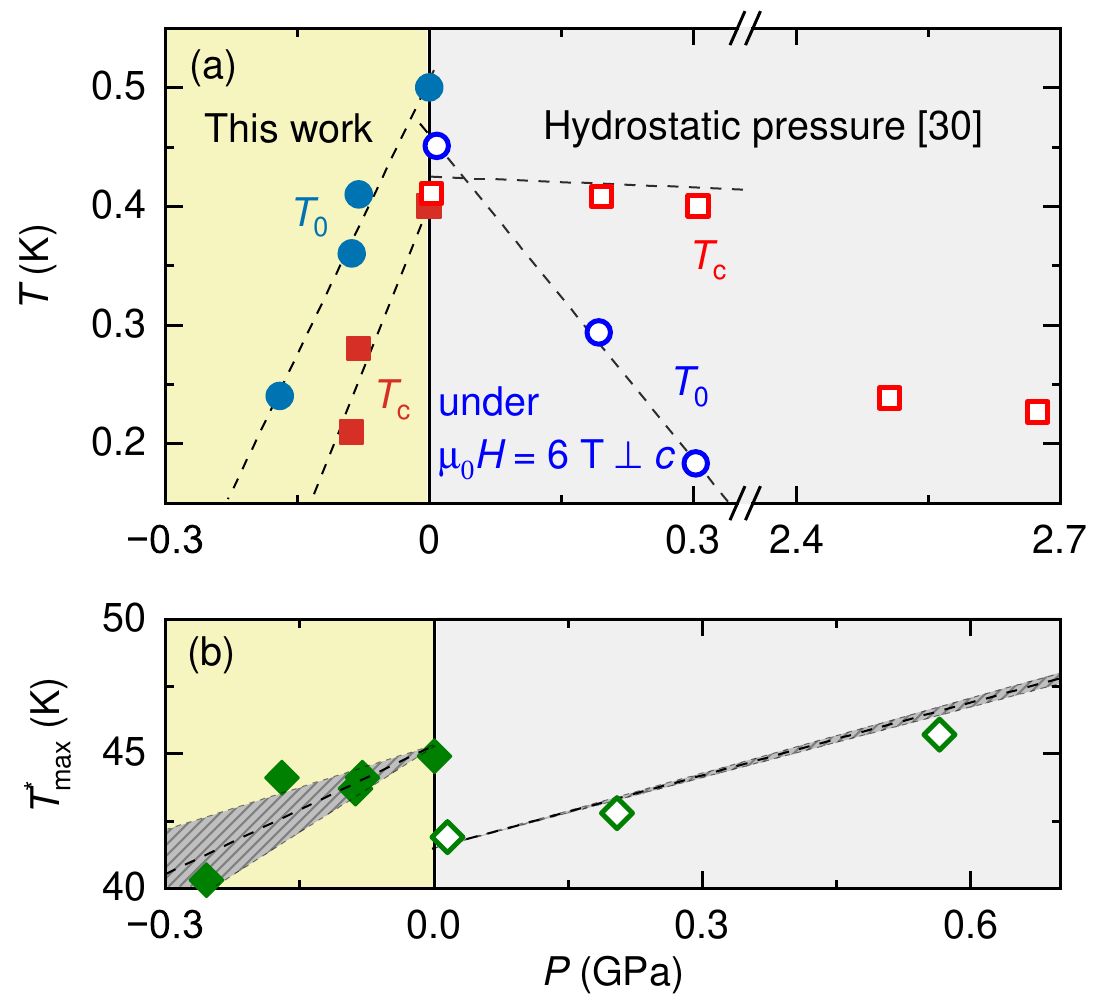}
    \caption{Phase diagrams with respect to pressure for (a) $T_{\mathrm{c}}$ and $T_{\mathrm{0}}$ and (b) $T^{*}_{\textrm{max}}$. The grey shaded region in (b) represents the uncertainty in the rate of change of $T^{*}_{\textrm{max}}$ with respect to pressure. The $T_{\mathrm{0}}$ transition under hydrostatic pressure \cite{Pfeiffer2024} was determined under an applied in-plane magnetic field of 6 T to observe the transition temperature by suppressing superconductivity. Since the $T_{\mathrm{0}}$ transition is insensitive to in-plane magnetic field in this field scale, this result should represent the zero-field properties of $T_{\mathrm{0}}$.}
    \label{fig:combined_phase_diagram}
\end{figure}

\section{\label{sec:level1}Discussion}

The main experimental findings can be summarized as follows:
(1) Partial substitution of Ce atoms with La induces a systematic lattice expansion in CeRh$_2$As$_2$.
(2) Simultaneously, it introduces additional disorder.
(3) The local moment of the Ce$^{3+}$ (4$f^1$)  remains intact up to $x$ = 0.1, together with persisting non-Fermi-liquid behavior as observed in pristine CeRh$_2$As$_2$.
(4) While $T^{*}_{\textrm{max}}$ decreases moderately, both $T_{\mathrm{c}}$ and $T_{\mathrm{0}}$ 
are rapidly suppressed with increasing $x$.
The observation (3) indicates that the La substitution can be considered as a perturbative tuning parameter rather than a fundamental change to the electronic structure.
This will allow us to interpret the changes in the low-temperature properties (4) as combined consequences of two effects: (1) the effective negative pressure from the lattice expansion and (2) additional disorder induced by chemical substitution.

To disentangle the effect of the lattice expansion from that of additional disorder on the transition temperatures of $T_{\mathrm{c}}$, $T_{\mathrm{0}}$, and $T^{*}_{\textrm{max}}$, we compare our results with previous hydrostatic-pressure studies \cite{Pfeiffer2024}, treating La substitution as an extension to the negative pressure regime.
Neglecting the slightly anisotropic lattice expansion, the effective hydrostatic pressure for each La concentration can be estimated from the change in the unit-cell volume based on the theoretical value of the bulk modulus, $B$ = 168 GPa \cite{ALI2023415224}.
This estimates the unit-cell expansion for the $x$ = 0.1 substitution (\textit{$\Delta V_\mathrm{unit-cell}$} $\approx$ 0.15 \%) to correspond to about -0.25 GPa.
The combined phase diagrams are shown in Fig. \ref{fig:combined_phase_diagram}.

As shown in Fig. \ref{fig:combined_phase_diagram}(b), $T^{*}_{\textrm{max}}$ continuously increases as the lattice shrinks in the overall pressure range. The dashed line on the right panel represents the guide passing through the pressure-study data measured upto ~ 2.69 GPa  \cite{Pfeiffer2024}.
This behavior is consistent with the expectation that a compression of the Kondo lattice strengthens the Kondo exchange interaction.
The discrepancy of the $T^{*}_{\textrm{max}}$ values at zero pressure may arise from a slight quality difference between the two independent samples used in the experiments.
The rate of the change of $T^{*}_{\textrm{max}}$ can be larger on the negative side (15.9 $\pm$ 5.5 K/GPa) than on the positive side (9.0 $\pm$ 0.3 K/GPa).
Although the size of the slope difference is still within the experimental error, the possible stronger suppression of $T^{*}_{\mathrm{max}}$ for negative pressures can be attributed to additional disorder effects.

On the other hand, $T_{\mathrm{c}}$ and $T_{\mathrm{0}}$, shown in Fig. \ref{fig:combined_phase_diagram}(a), demonstrate a contrasting pressure-dependent behavior on the positive and negative side.
While $T_{\mathrm{c}}$ quickly vanishes and is no longer resolvable below $\sim$ -0.2 GPa, it remains relatively robust under positive pressure.
In contrast, $T_{\mathrm{0}}$ is suppressed over a narrow range of pressure in both regimes.
Given that $T_{\mathrm{c}}$ and $T_{\mathrm{0}}$ are expected to be enhanced by increasing effective negative pressure, based on their behavior found under positive pressure, the observed suppression upon substitution cannot be explained by lattice expansion alone. 
These findings indicate that the primary origin of the changes in $T_{\mathrm{c}}$ and $T_{\mathrm{0}}$ is disorder introduced by substitution, rather than the hydrostatic pressure effect.

\begin{figure}[h]
    \centering
    \includegraphics[width=1\linewidth]{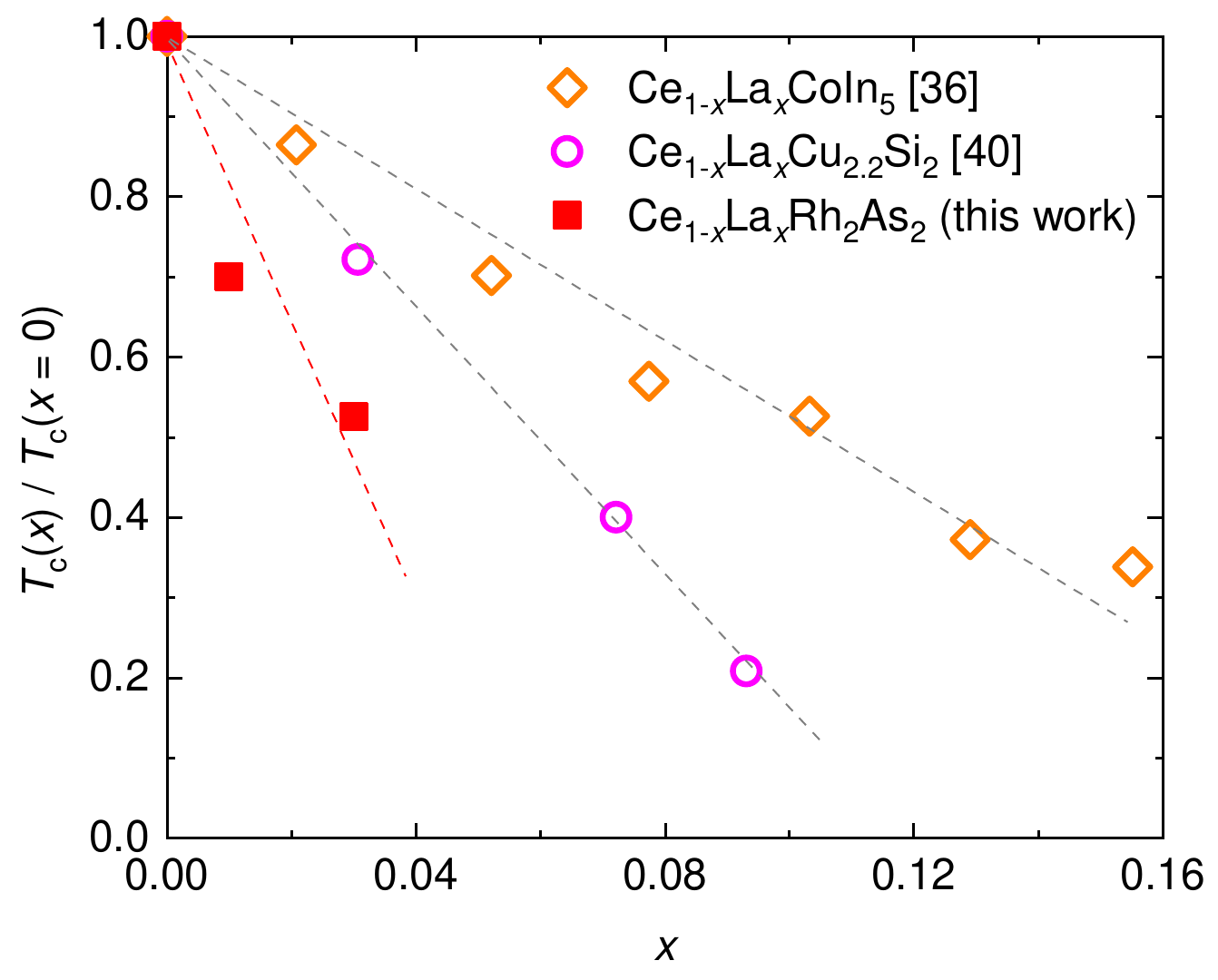}
    \caption{Evolution of the normalized $T_{\mathrm{c}}$,  $T_{\mathrm{c}}$/$T_{\mathrm{c}}$($x$ = 0), with respect to La concentration, $x$, in CeCoIn$_5$ \cite{Petrovic2002}, CeCu$_{2.2}$Si$_2$ \cite{AHLHEIM1988}, and CeRh$_2$As$_2$ (this work).}
    \label{fig:Ce_Tc_summary}
\end{figure}

The rapid suppression of $T_{\mathrm{c}}$ in CeRh$_2$As$_2$ is indicative of unconventional superconducting pairing.
Heavy-fermion superconductivity mediated by magnetic fluctuations is typically characterized by strongly anisotropic gap structures with a sign change of the order parameter, rendering it to be highly sensitive to impurity scattering.
In Fig. \ref{fig:Ce_Tc_summary}, we compare the normalized $T_{\mathrm{c}}$ under La substitution of CeRh$_2$As$_2$ with other Ce-based heavy-fermion superconductors, CeCoIn$_5$ \cite{Petrovic2002} and CeCu$_{2.2}$Si$_2$ \cite{AHLHEIM1988}.
Overall, La substitution reduces $T_{\mathrm{c}}$ by at least a factor of 2  at $x = 0.1$.
The differences in the relative $T_{\mathrm{c}}$ reduction reflect variations in underlying microscopic parameters \cite{STEGLICH1987}. While the Abrikosov-Gorkov model applied to our experimental data might associate the reduction of $T_c$ with an anisotropic gap symmetry \cite{AG_1960}\cite{Openov_1998}, we could not draw a reasonable conclusion. The detailed analysis are provided in the supplemental material \cite{SM}.
Nevertheless, superconductivity in CeRh$_2$As$_2$ appears to be the most sensitive among the three systems \cite{SM}.
Here, we consider the unique structural character of CeRh$_2$As$_2$ to be potentially responsible for the particular vulnerable nature of $T_{\mathrm{c}}$.
As mentioned earlier, the lack of inversion symmetry on the Ce site introduces sublattice degrees of freedom that are involved in the SC order parameter \cite{Fischer2023}.
The requirement for such spatial pair-wise coherence which is essential for maintaining superconductivity could make the system more susceptible to local disorder than the other Ce-based superconductors.
This additional pair-breaking mechanism will be independent from the one given by an anisotropic gap structure, which may explain the more rapid $T_{\mathrm{c}}$ suppression upon La substitution in CeRh$_2$As$_2$.

Finally, we turn to the evolution of $T_{\mathrm{0}}$ under hydrostatic pressure and La substitution.
Since hydrostatic pressure suppresses $T_{\mathrm{0}}$ upon lattice compression, one could naively expect that the negative effective pressure given by La substitution could instead enhance $T_{\mathrm{0}}$.
However, our result shows a decrease in $T_{\mathrm{0}}$ upon lattice expansion, even at a similar rate in terms of effective pressure.
This suggests that the $T_{\mathrm{0}}$ suppression by La substitution is primarily caused by a disorder effect.
The destabilization of phase I upon both disorder and lattice change may support the proposed itinerant origin of the phase \cite{Hafner2022,TChen2024,Wu2024,Chen2024,XChen2024}, driven by Fermi-surface nesting.
In such a scenario, the nesting condition is highly sensitive to a change in the Fermi-surface geometry (by hydrostatic pressure) and a reduction of coherent scattering between the nested part of the Fermi surface (by substitution).

To summarize, we have investigated La-substituted CeRh$_2$As$_2$ single crystals up to $x$ = 0.1.
Substitution of Ce with La induces a systematic increase of the $a$ and $c$ lattice parameters without significantly altering the local Ce$^{3+}$ valence state.
The resulting negative effective pressure is reflected in a gradual decrease of  the local resistivity maximum, $T^{*}_{\mathrm{max}}$, consistent with the reduction of the Kondo-energy scale upon lattice expansion. 
However, the rapid suppression of both the SC and phase I is attributed to the dominant role of disorder induced by substitution.
These findings can be related to the unconventional pairing nature of superconductivity and the itinerant origin of phase I.

\section{Acknowledgments}
S.L.R.S. and S.K. were supported by the Max Planck Society and the
Deutsche Forschungsgemeinschaft (DFG) - KH 387/1-1. J. W. acknowledges   support from DFG through the W\"{u}rzburg-Dresden Cluster of Excellence on Complexity, Topology and Dynamics in Quantum Matter -- $ctd.qmat$ (EXC 2147, Project No.\ 390858490), as well as the support of the HLD at HZDR, member of the European Magnetic Field Laboratory (EMFL).

\section{Data Availability}
The data that support the findings of this article are openly available at \cite{raw_data}.

\nocite{*}
\bibliography{references}
\end{document}


\title{Supplementary material for La substitution studies on the heavy-fermion superconductor CeRh$_2$As$_2$}

\author{Sushma Lakshmi Ravi Sankar}
\affiliation{Max Planck Institute for Chemical Physics of Solids, N\"othnitzer Stra{\ss}e 40, 01187 Dresden, Germany}
\affiliation{Institute for Solid State and Materials Physics, Technische Universit\"at Dresden, 01062 Dresden, Germany}

\author{Manuel Brando}
\affiliation{Max Planck Institute for Chemical Physics of Solids, N\"othnitzer Stra{\ss}e 40, 01187 Dresden, Germany}

\author{Jochen Wosnitza}
\affiliation{Institute for Solid State and Materials Physics, Technische Universit\"at Dresden, 01062 Dresden, Germany}
\affiliation{Dresden High Magnetic Field Laboratory (HLD-EMFL) and W\"urzburg-Dresden Cluster of Excellence ctd.qmat, Helmholtz-Zentrum Dresden-Rossendorf, 01328 Dresden, Germany}

\author{Seunghyun Khim}
\email[Corresponding author: ]{Seunghyun.Khim@cpfs.mpg.de}
\affiliation{Max Planck Institute for Chemical Physics of Solids, N\"othnitzer Stra{\ss}e 40, 01187 Dresden, Germany}

\date{\today}

\maketitle

\section{Determination of $T_0$}

\begin{figure}[h]
    \centering
    \includegraphics[width=1\linewidth]{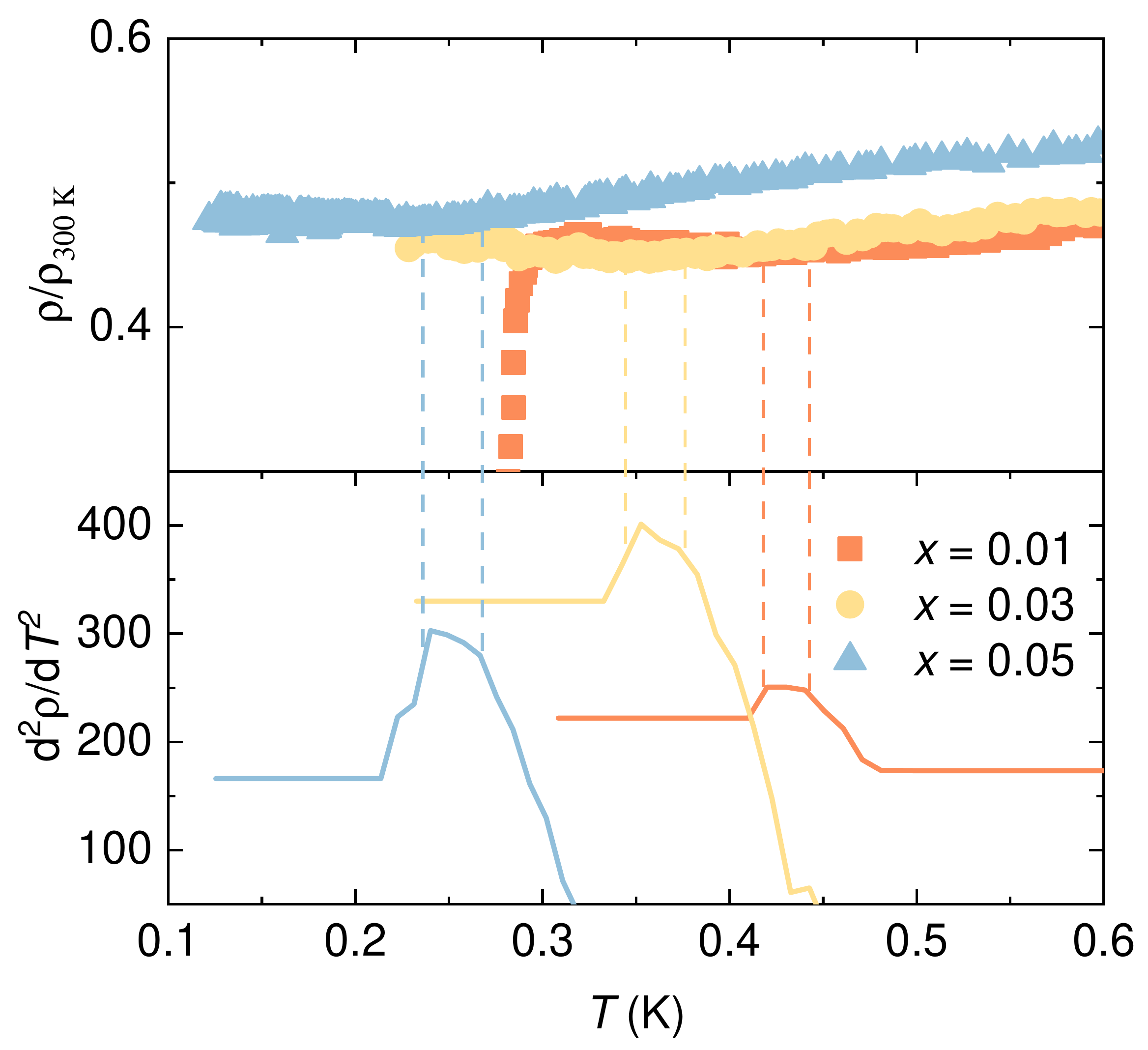}
    \caption{Plot of normalized resistivity $\rho$/$\rho_{\mathrm{300 K}}$  vs. temperature for different La concentration $x$ is presented in the upper panel and the second derivative of the data is shown in the lower panel to depict the decrease in Phase I transition temperature, $T_0$ with $x$. }
    \label{fig:T0}
\end{figure}

The transition into phase I at $T_0$ appears as an upturn in resistivity. 
We defined $T_0$ at the temperature where the second derivative of resistivity, $d^{2}\rho / dT^2$, is maximized. This refers to the point of maximum curvature in the resistivity upturn.
In the analysis, the data is first reduced to evenly spaced temperature values with 10 mK as sampling interval and Y-mean (mean of the resistivity values in the interval) as the sampling method. 
The second derivative operation is then performed with Savitzky-Golay smoothening. Fig. \ref{fig:T0} shows the normalized resistivity values for different $x$ along with the $d^{2}\rho / dT^2$ curves.

\section{Estimation of the concentration of isolated paramagnetic spins induced by 10 \% L\lowercase{a} substitution}

\begin{figure}[h]
    \centering
    \includegraphics[width=1\linewidth]{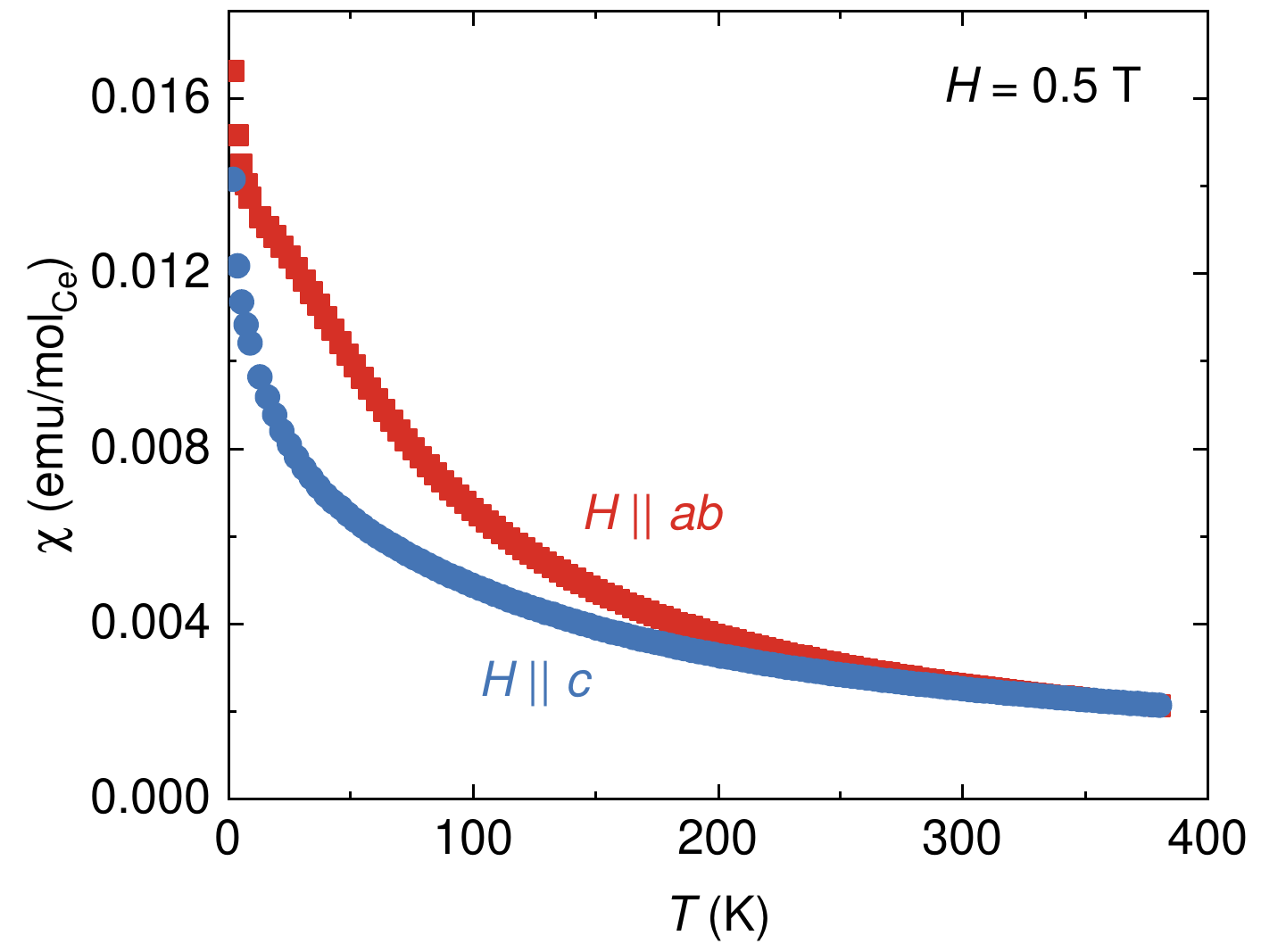}
    \caption{Susceptibility against temperature plot for $x$ = 0.10 along $ab$-plane (red squares) and $c$-axis (blue circles) showing diverging behavior at low temperature.}
    \label{fig:chi_T}
\end{figure}

\begin{figure}[h]
    \centering
    \includegraphics[width=1\linewidth]{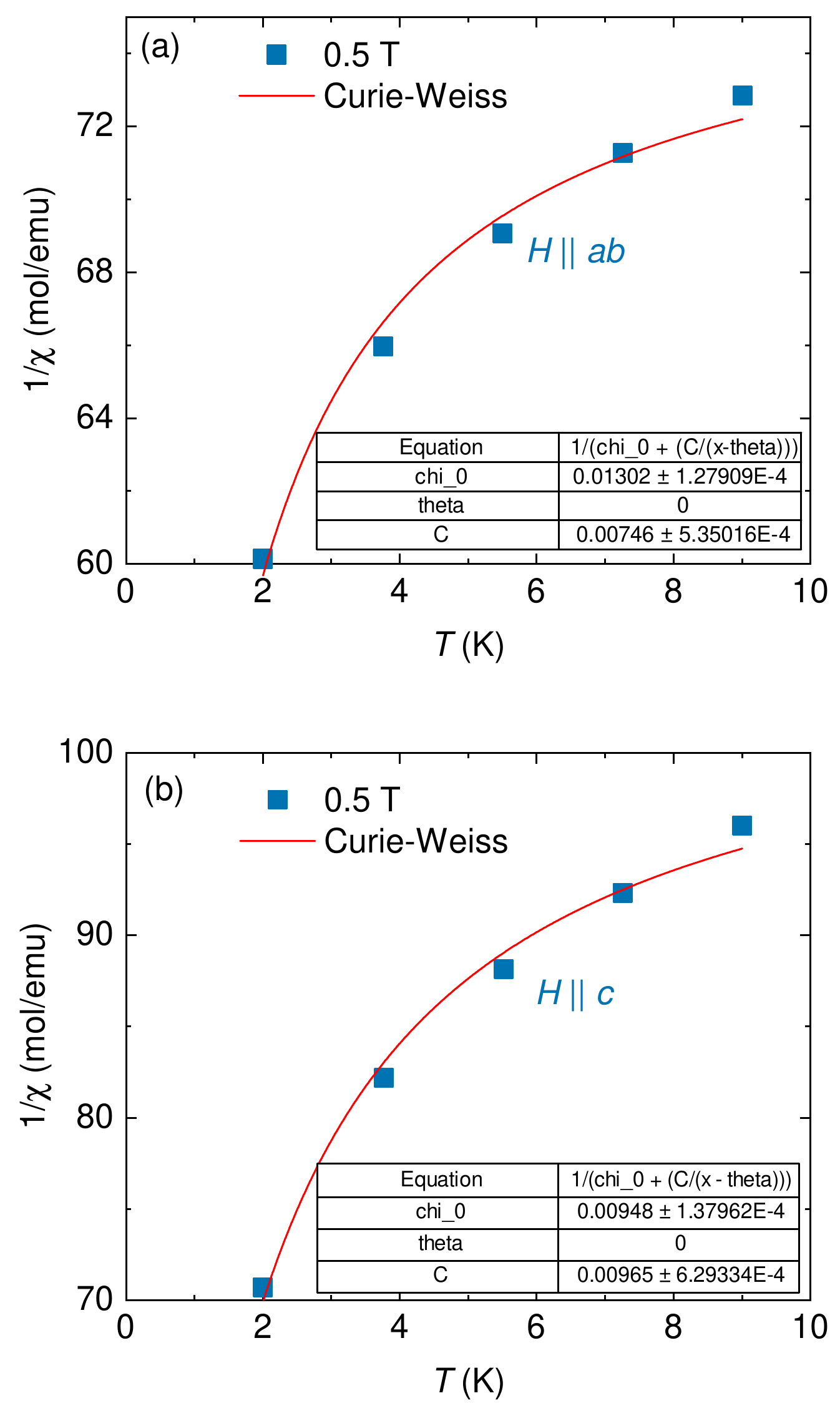}
    \caption{The plot of inverse susceptibility, 1/$\chi$ against temperature for $x$ = 0.10 with magnetic field (a) along ${ab}$-plane and (b) along $c$-axis. The data selected corresponds to the temperature range where we observe the Curie tail. }
    \label{fig:1_chi_T}
\end{figure}

The temperature-dependent susceptibility data for $x$ = 0.10 shows diverging behavior below 10 K as shown in Fig. \ref{fig:chi_T}.
Since the Ce$^{3+}$ and La$^{3+}$ moments are effectively nonmagnetic in this temperature range due to the Kondo screening and the lack of 4$f$ electron, respectively, the observed Curie-Weiss tail is attributed extrinsic origins, such as structural defects and paramagnetic impurities introduced by substitution.
Based on the assumption that the diverging behavior arises from isolated spin-1/2 moments with a gyromagnetic factor $g$ = 2, we applied the Curie-Weiss model to fit the inverse-susceptibility data below 10 K. 
The fitting curves are shown in Fig. \ref{fig:1_chi_T}. 
The obtained Curie constants correspond to an effective concentration of the paramagnetic moments equivalent to 14 \% and 16 \% of the Ce/La site for the in-plane and $c$-axis susceptibilities, respectively. 
These values are comparable to the nominal La concentration (10 \%), suggesting that the assumed condition is physically reasonable.

\section{Abrikosov-Gorkov (AG) formalism fit}

The AG Cooper pair-breaking theory originally formulated for $T_c$ suppression by magnetic impurity scattering in conventional SC \cite{AG_1960} is extended for non-magnetic scatterers in an effectively single-band unconventional SC. 
The formula is represented as \cite{Openov_1998}:

\begin{equation}
\ln\frac{T_{c0}}{T_c}
=
\chi \left[
\psi\left(\frac{1}{2} + \frac{\hbar/\tau}{4\pi k_B T_c}\right)
-
\psi\left(\frac{1}{2}\right)
\right]
\tag{1}
\end{equation}

where $\psi$ is the digamma function, $\hbar$ is Planck’s constant, 1/$\tau$ is the quasi particle scattering rate, $k_B$ is Boltzmann’s constant, $T_{c0}$ is the superconducting transition temperature of the pristine system, and 

\begin{equation}
\chi = 1 - \frac{\langle \Delta(\mathbf{k}) \rangle_{FS}^{2}}{\langle \Delta^{2}(\mathbf{k}) \rangle_{FS}}
\tag{2}
\end{equation}

represents the anisotropy of the superconducting order parameter averaged over the Fermi surface. 
The pair-breaking energy scale, 1/$\tau$, is related to the change in in-plane residual resistivity, $\rho_0$, using the relation $\frac{1}{\tau} = \rho_0 (\frac{\omega_{pl}^{2}}{4\pi})$, which follows from Drude model \cite{ashcroft_mermin_1976}. 
Here, $\hbar\omega_{pl}$ is the characteristic plasma frequency of the in-plane charge carriers. 
The corresponding modified AG fitting equation is \cite{Radtke_Levin_1993}\cite{Ranna_2025}:

\begin{equation}
\ln\frac{T_{c0}}{T_c}
=
\chi \left[
\psi\left(\frac{1}{2} + \frac{C\Delta\rho_0}{T_c}\right)
-
\psi\left(\frac{1}{2}\right)
\right]
\tag{3}
\end{equation}

where $C = \frac{\omega_{pl}^{2}\epsilon_0\hbar}{4\pi k_B}$ is a scaling parameter with dimensions of kelvin per resistivity. 
The fitting equation (3) has two free parameters, $\chi$ and $C$ which can be adjustable to fit the data.

\begin{figure}
    \centering
    \includegraphics[width=1\linewidth]{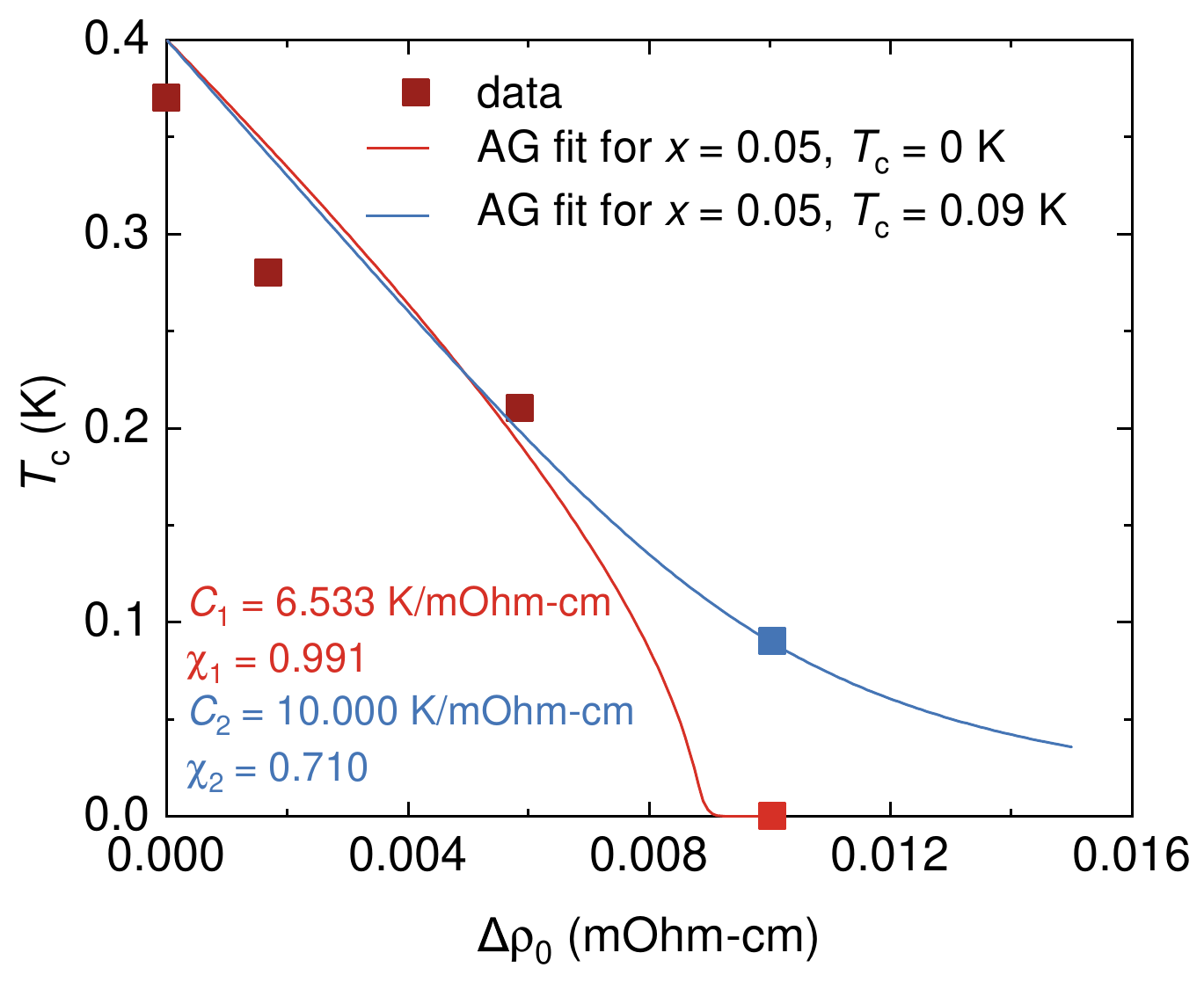}
    \caption{$T_c$ as a function of the change in the residual resistivity $\Delta \rho_0$ which is given from the relation of Eq. R1. The solid lines represent AG fit using Eq. 3, where $\chi$ and C are free parameters. The red and blue lines represent the fit on two extreme cases, $T_c$ = 0 and $T_c$ = 0.09 for $x$ = 0.05, respectively.}
    \label{fig:AG}
\end{figure}

In our study, the AG fitting was not straight forward due to: (1) the lack of information on the critical La concentration that completely suppresses $T_c$ and (2) highly scattered relation between $T_c$ and the absolute value of the residual resistivity, $\rho_0$. 
Therefore, we proceed with the analysis with some assumptions as following:

    \begin{enumerate}
        \item $T_c$ of the undoped CeRh$_2$As$_2$ is an intrinsic maximum, $T_{c0}$, without impurity scattering.
        \item 	The issue (2) might be given by experimental error in determining the accurate resistivity value. We alternatively estimated the residual resistivity for 0 $\leq x \leq$ 0.1 assuming a linear relation between the absolute residual resistivity, $\rho_0$ and normalized residual resistivity, $\rho_n$ = $\rho_0/\rho_{300 K}$ = 1/RRR as follows: 
        \begin{equation}
            \rho_0(x) = \rho_0(x=0)\cdot\frac{\rho_n(x)}{\rho_n(x=0)}
            \tag{4}
        \end{equation}
        Since the absolute residual resistivity of CeRh$_2$As$_2$, $\rho_0$($x$ = 0), is known, $\rho_0$($x$) in the range $0 \leq x \leq 0.1$ can be estimated.
        \item Regarding the issue (1), we considered two extreme cases: [Case 1] $T_c$ is completely suppressed to $T_c$ = 0 K for $x$ = 0.05 and [Case 2] $T_c$ is 0.09 K (considering a transition width of 0.04 K), just below the lowest temperature (0.12 K) the measurement setup can reach.
    \end{enumerate}

The two fitting curves are shown in Fig. \ref{fig:AG}. 
$\chi_1$ and $C_1$ are the parameters corresponding to [Case 1] $T_c$ = 0 K for $x$ = 0.05 and $\chi_2$ and $C_2$ are for [Case 2] $T_c$ = 0.09 K, $x$ = 0.05. 
The results of the fits on two extreme cases propose that the gap anisotropy parameters can be in the range of 0.71 $\le$ $\chi$ $\le$ 0.99.
The obtained $C$ values correspond to 0.23 eV $\leq \hbar\omega_{pl} \leq$ 0.28 eV. 
The range of the plasma frequency, $\hbar\omega_{pl}$, is two orders of magnitude smaller than the value of 31.7 eV, proposed by a theoretical work \cite{ALI2023415224}. 
However, this value is not completely away from the the values discussed in other heavy-fermion systems, $\hbar\omega_{pl}$ = 0.28 eV for UPt$_3$ \cite{MARABELLI1986287} and 0.15 eV for CeCu$_6$ \cite{Marabelli_1990}, implying that our result might not be completely unrealistic.

While the fit cannot be considered to be quantitative due to the inherent limit on the experimental data, they seem to be phenomenologically informative, proposing unconventional pair breaking due to an anisotropic superconducting gap.

\section{Crystal electric field analysis}

\begin{figure}
    \centering
    \includegraphics[width=1\linewidth]{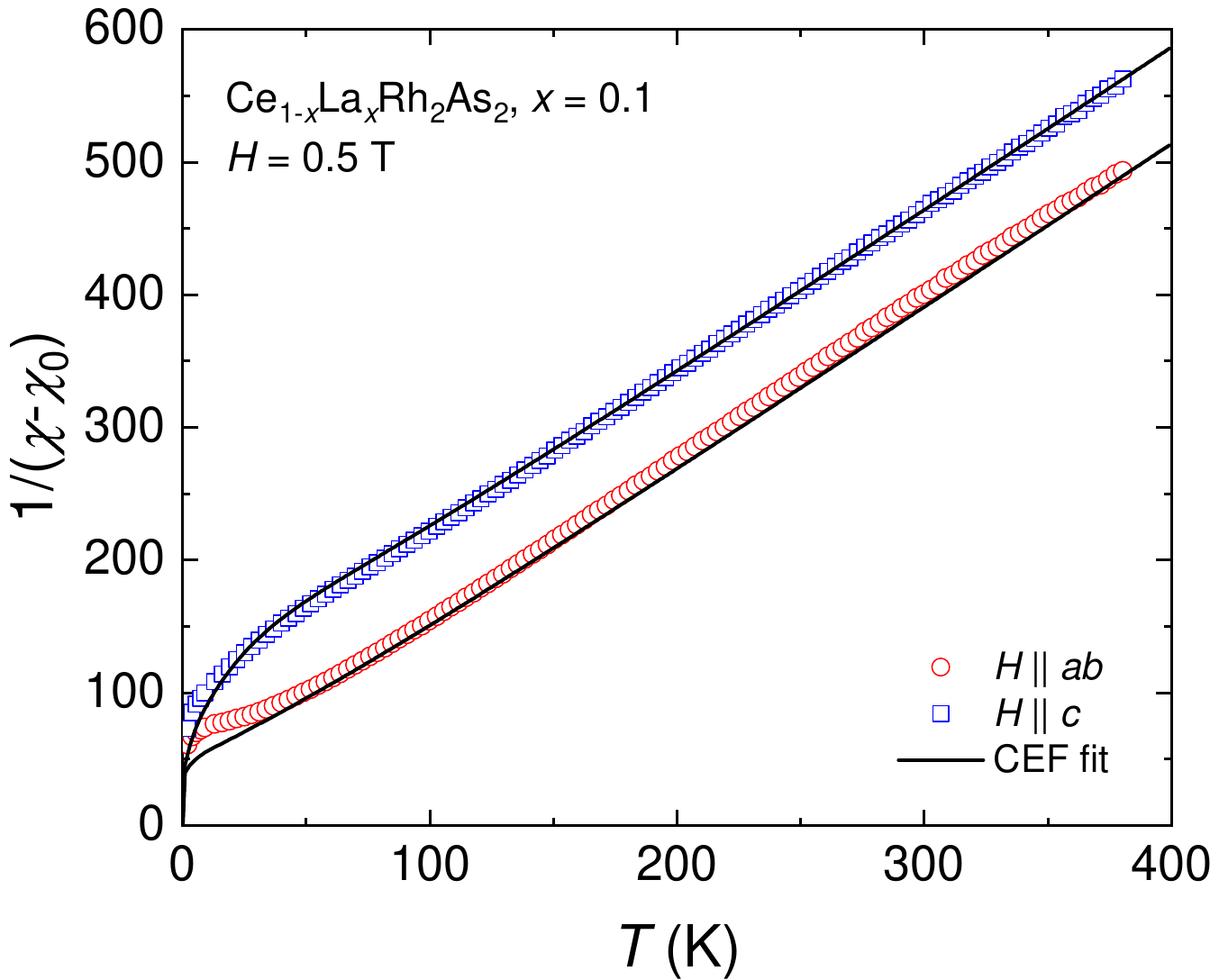}
    \caption{Temperature dependence of inverse magnetic susceptibility  for $H$ ‖ $ab$ (red circles) and $H$ ‖ $c$ (blue squares), respectively. $\chi_0$ are 9.51 $\times$ 10${-5}$ emu/mol and 3.47 $\times$ 10${-4}$ emu/mol for $H \parallel ab$ and $H \parallel c$, respectively. The solid lines correspond to the CEF fit on each data.}
    \label{fig:CEF}
\end{figure}

The crystal field Hamiltonian for Ce in the tetragonal environment is given by $H = B_2^0 O_2^0 + B_4^0 O_4^0 + B_4^4 O_4^4$ where $B_n^m$ are the crystal electric field parameters and $O_n^m$ are Stevens operators \cite{HUTCHINGS1964227}\cite{Stevens_1952}. 
The parameter $B_2^0$ can be estimated from the Weiss temperatures using the relation, $B_2^0 = (\theta_W^{ab} - \theta_W^c)\cdot\frac{10k_B}{3(2J-1)(2J+3)}$. 
The magnetic susceptibility is calculated as 

\begin{widetext}
\begin{equation}
\chi_{\mathrm{CEF},i}
=
N_A (g_J \mu_B)^2 \frac{1}{Z}
\left(
\sum_{m \ne n}
2 \left| \langle m \lvert J_i \rvert n \rangle \right|^2
\frac{1 - e^{-\beta (E_n - E_m)}}{E_n - E_m}
e^{-\beta E_n}
+
\sum_n
\left| \langle n \lvert J_i \rvert n \rangle \right|^2
\beta e^{-\beta E_n}
\right)
\tag{4}
\end{equation}
\end{widetext}

where $Z=\sum_n e^{(-\beta E_n)}, \beta=\frac{1}{k_B T}$, and $i=x,y,z$. The calculated inverse magnetic susceptibility including the molecular field contribution $\lambda_i$ as $\chi_i^{-1}=\chi_{CEF,i}^{-1} - \lambda_i$ is given to fit the experimental data. 

From the high-temperature Curie-Weiss fit, the Weiss temperatures for $x$ = 0.10 (La 10 $\%$) are determined to be $\theta_W^{ab}$ = 24.8 K and $\theta_W^c$ = 82.8 K. 
The calculated $B_2^0$ parameter is 6.04 K, which is slightly smaller than $B_2^0$ = 6.5 K for the pristine CeRh$_2$As$_2$. 
By adopting the same crystal electric field parameters, $B_4^0$ = 0.1 K and $B_4^4$ = 2.8 K, as determined for CeRh$_2$As$_2$, the CEF states were calculated by the diagonalization of the crystal field Hamiltonian. 
The obtained states are a ground state doublet $|\Gamma_7^{(1)}\rangle = - 0.88 |\mp3/2\rangle + 0.48 |\pm5/2\rangle$ with  the first excited state $|\Gamma_6\rangle = |\pm1/2\rangle$ at 33.9 K and the second excited level $|\Gamma_7^{(2)}\rangle = 0.48 |\pm3/2\rangle + 0.88 |\mp5/2\rangle$ at 176.76 K. 
The calculated susceptibility fit (shown in Fig. \ref{fig:CEF}) with $\lambda_i$ set to 36 mol/emu reproduced the experimental data well.

\bibliography{references}